\documentclass[]{SSRv}

\usepackage{graphicx}
\usepackage{mathptm}
\usepackage{url}

\begin{document}
\def\ms{M$_{\odot}$}
\begin{article}
\begin{opening}

\title{Origin and evolution of the light nuclides}
\runningtitle{Origin and evolution of the light nuclides  }

\author{N. \surname{Prantzos}}
\institute{
 Institut d'Astrophysique de Paris, 98bis Bd Arago, 75014 Paris\\
(Email: prantzos@iap.fr)}

\runningauthor{N. Prantzos}

\received{10 January 2007} \revised {} \accepted{2007}

\begin{abstract}
After a short historical (and highly subjective) introduction to the field, I discuss our current understanding of the origin and evolution of the light nuclides D, $^3$He, $^4$He, $^6$Li, $^7$Li, $^9$Be, $^{10}$B and $^{11}$B.
Despite considerable observational and theoretical progress,  important uncertainties still persist for each and every one of those nuclides. The present-day abundance of D in the local interstellar medium is currently uncertain, making it difficult to infer the recent  chemical evolution of the solar neighborhood. To account for the observed quasi-constancy of $^3$He abundance from the Big Bang to our days, the stellar production of that nuclide {\it must} be negligible; however, the scarce observations of its abundance in planetary nebulae  seem to contradict this idea. The observed Be and B evolution as primaries suggests that the source composition of cosmic rays has remained $\sim$constant since the early days of the Galaxy, a suggestion with far reaching implications for the origin of cosmic rays; however, the main idea proposed to account for that constancy, namely that superbubbles are at the source of cosmic rays, encounters some serious difficulties. The best explanation for the mismatch between primordial Li and the observed "Spite-plateau"  in halo stars appears to be depletion of Li in stellar envelopes, by some yet poorly understood mechanism. But this explanation impacts on the level  of the recently discovered early ``$^6$Li plateau'', which (if confirmed), seriously challenges current ideas of cosmic ray nucleosynthesis. 
\end{abstract}


\end{opening}

\section{Introduction}

In their monumental study on ``Synthesis of the Elements in Stars'', Burbidge et al. (1957, B$^2$FH) recognized the difficulty of finding a nuclear process able to synthesize the light nuclides D, $^6$Li and $^7$Li, $^9$Be,  $^{10}$B and $^{11}$B. Indeed, these nuclides are so fragile (as revealed by their binding energies in Fig. 1) that they are consummed in stellar interiors, once hydrogen-rich material  is brought to temperatures higher than 0.6 MK for D, 2 MK for $^6$Li, 2.5 MK for $^7$Li, 3.5 MK for $^9$Be and  5 MK for the boron isotopes\footnote{In such temperatures, and for densities comparable to those encountered in the bottom of the outer convective zones of low mass stars, like the Sun, the lifetimes of light nuclides against proton captures are smaller than a few  Gyr.}.

B$^2$FH argued that the ``x-process'' (as they called the unknown nucleosynthetic mechanism) should occur in low-density, low-temperature environments. They discussed stellar atmospheres (of active, magnetized stars) and gaseous nebulae (traversed by energetic particles) as possible sites, and they concluded that, most probably, D originates from a different process than the  Li, Be and B (hereafter LiBeB) isotopes.

The synthesis of the He isotopes ($^3$He and $^4$He) drew very little attention in B$^2$FH, where it was flatly attributed to stellar H-burning with no further comments. This (most suprising) neglect of  B$^2$FH was corrected in Hoyle and Tayler (1964), who demonstrated that H-burning stars of the Milky Way (MW), releasing an energy of $\varepsilon(H\rightarrow^4He)$=6 10$^{18}$ erg g$^{-1}$, having a total mass $M_{MW}$=10$^{11}$ \ms \  and shining collectively with a luminosity $L_{MW}$=6 10$^{43}$ erg s$^{-1}$  for $T$=10$^{10}$ yr, could produce a mass fraction of $^4$He of only a few per cent; this is  about 10 times less than the observed abundance of $^4$He (mass fraction X($^4$He)$\sim$0.25), which requires then another nucleosynthesis site, like the  hot early universe (or, in Hoyle's views, high temperature explosions of extremely massive pre-galactic stars).

\begin{figure*}
\centering
\includegraphics[clip=,angle=-90,width=0.9\textwidth]{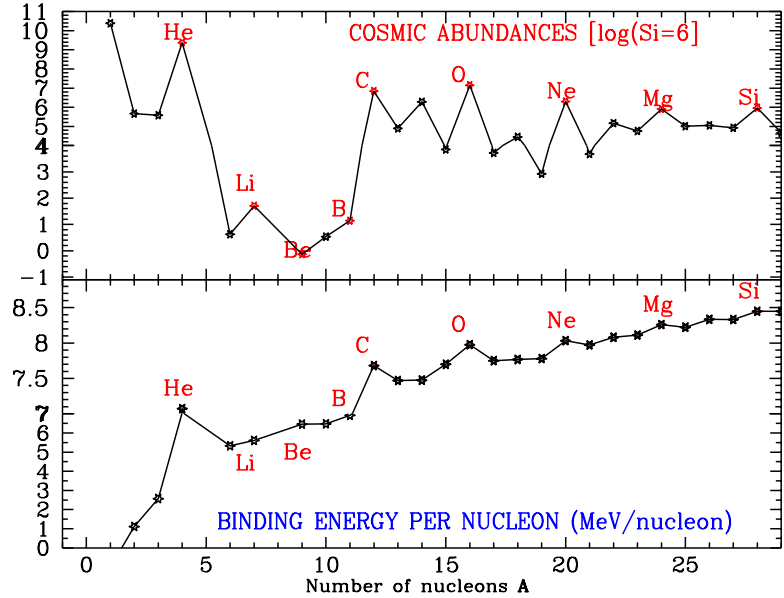}
\caption{{\it Top:} Cosmic abundances of the light nuclides, 
from H to Si, in the log(Si)=6 scale. Symbols indicate major isotopes of a given element; alpha-nuclides (C-12, O-16, etc.) dominate always their neighbors, up to Ca-40. 
{\it Bottom:} Binding energies of
the light nuclides (note the change in the vertical scale at 7 MeV/nucleon). 
D, $^3$He, and LiBeB isotopes are more fragile than neighboring nuclei; that fragility is clearly reflected in the cosmic abundance curve.} 
\label{jdrfig1}
\end{figure*}

After the discovery of the cosmic microwave background (CMB) by Penzias and Wilson (1965), which strongly supported the Big Bang model for the origin of the Universe, calculations of primordial nucleosynthesis by Peebles (1966) and Wagoner et al. (1967) showed that D, $^3$He and $^4$He could be produced in large amounts (i.e. comparable to their present-day measured values) in the hot early Universe. Moreover, the latter work showed that significant amounts of $^7$Li could also be generated in that rapidly cooling environment.

\begin{figure*}
\centering
\includegraphics[clip=,width=0.8\textwidth]{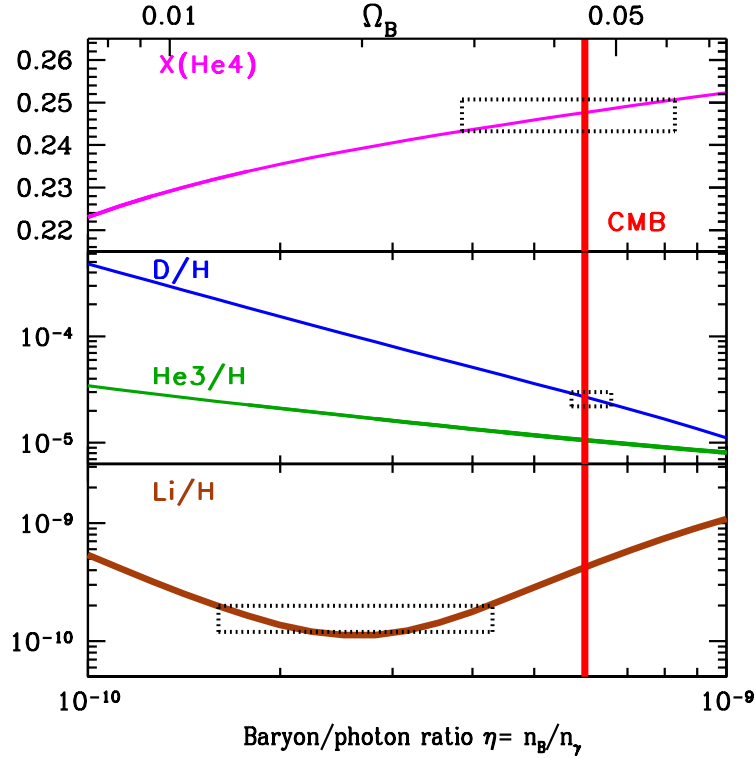}
\caption{Results of standard Big Bang nucleosynthesis (SBBN) calculations vs. 
cosmic baryon/photon ratio $\eta$ ({\it bottom horizontal axis}) or cosmic baryon density (in units of the critical density, {\it top horizontal axis}). A Hubble constant of H$_{0}$=70 km/s/Mpc is assumed. The width of the curves reflects 1-$\sigma$ statistical uncertainties in the 12 main nuclear reaction rates of SBBN, still important in the case of $^7$Li. Abundances are by number, except for $^4$He, for which the mass fraction is given. The {\it vertical line} across all panels corresponds to the baryonic density determined by analysis of the Cosmic Microwave Background anisotropies, detected by WMAP. {\it Boxes} indicate observed, or observationally inferred, primordial abundances of the light nuclides (see text). The agreeement with WMAP is extremely good in the case of D and poor in the case of $^7$Li (a factor of 2-3), while determination of primordial $^4$He suffers from considerable systematic uncertainties.} 
\label{jdrfig2}
\end{figure*}

In the early 1970ies, the pre-solar abundances of D and $^3$He were accurately established (Black 1971, Geiss and Reeves 1972) and it was convincingly argued that D could be produced in no realistic astrophysical site other than the hot early Universe  (Reeves et al. 1973, Epstein 1976). Since D can only be destroyed by astration after the Big Bang, it can then be used as a ``baryometer'' (Reeeves et al. 1973), revealing that  the cosmic baryonic density is smaller than the critical value (e.g. Gott et al. 1974), i.e. that baryons cannot "close" the Universe. The precise value of the baryonic density was pinpointed only 30 years later, from converging measurements of the CMB anisotropies by the WMAP satellite and observations of (presumably primordial) D abundances in remote gas clouds (Fig. 2 and texts by Tytler, Geiss, Reeves in this volume). However, the exact amount of D astration during galactic evolution remains unknown at present, due to uncertainties on its present-day local ISM value (see Sec. 2.1 and text by Linsky in this volume).

The case of $^3$He turned out to be much more complex than the one of D, since $^3$He can be produced not only in the Big Bang, but also in stars (from burning of primordial D: $D+p\longrightarrow~D$ {\it and} from the $p-p$ chains), whereas it may also be destroyed in stellar zones hotter than 10$^7$ K. In the first comprehensive Galactic chemical evolution model ever made (Truran and Cameron 1971), it was shown that, {\it if} standard stellar nucleosynthesis prescriptions are adopted for $^3$He (e.g. from models by Iben 1967), then that nuclide is largely overproduced during galactic evolution. Thirty six years later, the issue is not satisfactorily settled, despite theoretical and observational developments (see Sec. 2.2 and text by Bania in this volume).

The suggestion of B$^2$FH that substantial production of light nuclides can occur on the surfaces of active stars, further elaborated in Fowler et al. (1962), was refuted by Ryter et al. (1970) on the grounds of energetics arguments: the total amount of available (gravitational) energy, mostly  in the active T Tauri phase of stellar youth, is insufficient for that. Noting that the LiBeB/CNO ratio in galactic cosmic rays (GCR) is $\sim$10$^4$ times higher than in the ISM, Reeves et al. (1970) proposed that LiBeB isotopes are produced by spallation reactions on CNO nuclei, occuring during the propagation of GCR in the interstelar medium (ISM) of the Galaxy. The process was  definitely modeled by Meneguzzi et al. (1971, MAR), who found that the pre-solar abundances of $^6$Li, $^9$Be and $^{10}$B, $\sim$20\% of $^7$Li and $\sim$60\% of $^{11}$B can be produced that way, after 10 Gyr of galactic evolution. The majority of $^7$Li should originate in another (presumably stellar) site, unidentified as yet. AGB stars, where the Cameron-Fowler (1971) process may operate, appear as an attractive possibility (supported by observations of Li-rich evolved stars), but explosive H-burning in novae remains an interesting alternative.

\noindent
\begin{table}
\caption{Abundances of primordial nuclides ({\it from references in parenthesis})
\label{fig:Table1}}
\begin{center}
\begin{tabular}{ccccc}
\hline 
Nuclide           & SBBN+WMAP     & Observed earliest  & Pre-solar        & Local ISM  \\
                  & --13.7 Gyr      &    --(10-13) Gyr   &  --4.6 Gyr       &   Today     \\
 \hline
  D/H (10$^{-5}$)      & 2.56$\pm$0.18 {\it (1)} & 2.6$\pm$0.4  {\it (1)}   & 2.$\pm$0.35  {\it (2)} & 2.3$\pm$0.24 {\it (3)}\\
                       &                 &                    &                  & 0.98$\pm$0.19 {\it (4)}  \\
                       &                 &                    &                  &                          \\
  $^3$He/H (10$^{-5}$) & 1.04$\pm$0.04 {\it (1)} &                    & 1.6$\pm$0.06 {\it (2)} & 2.4$\pm$0.7 {\it 5)} \\
                      &                 &                    &       & 1.7$\pm$0.7 {\it (6)}  \\
                      &                 &                    &                  &                          \\
  $^4$He (Y$_P$)       & 0.2482$\pm$0.0007 {\it (1)} & 0.2472$\pm$0.0035 {\it (1)}   &  0.274 {\it (7)}          &             \\
                      &                 &                    &                  &                          \\
  $^7$Li/H (10$^{-10}$)& 4.44$\pm$0.57 {\it (1)}    & 1.1-2.  {\it (8)}   &  22.8 {\it (7)}    &             \\
                      &                 &                    &                  &                          \\
  $^6$Li/H (10$^{-10}$)& 0.0001    {\it (9)} &  0.08 {\it(10)}   &    1.73 {\it (7)}         &             \\
\hline 
\end{tabular}
\end{center}
\noindent {\footnotesize
{\it (1)}: Steigman (2006) and references therein (note the discussion on $Y_{P,OBS}$);
{\it (2)}: Geiss and Gloeckler (2002); {\it (3)}: Linsky et al. (2006); {\it (4)}: Hebrard et al. (2005);
{\it (5)}: Gloeckler and Geiss (1996); {\it (6)}: Salerno et al. (2003)}; 
{\it (7)}: Lodders (2003) ; {\it (8)}: Gratton (this volume) ;{\it (9)}: Serpico et al. (2004);
{\it (10)}: Asplund et al. (2006)
\end{table}

The major development of the 80ies was the discovery (Spite and Spite 1982) that the Li abundance in metal-poor halo stars remains constant, at about 0.05 of its pre-solar value (the "Spite plateau"). This behaviour, shared by no other metal, suggests that early  Li is primordial and gives further support to the theory of the Big Bang. However, the cosmic baryonic density inferred from WMAP measurements of CMB corresponds to a Li abundance 2-3 times higher than the Spite plateau (Fig. 2) and makes the statement "the Li plateau is primordial" sound rather strange. After a flurry of possible explanations, it appears now that the fault lies within the stars themselves (see Sec. 4.1 and text by Gratton, this volume), able to transform the primordial plateau into another, lower lying,  one.

Two major developments were made in the field in the 90ies. First, it was realized that the two LiBeB isotopes underproduced in the standard GCR scenario of MAR ($^{11}$B and - to a smaller extent - $^7$Li), can also be produced by neutrino-induced nucleosynthesis during core-collapse supernova explosions (Woosley et al. 1990); however, various uncertainties (neutrino spectra, structure of progenitor star, explosion mechanism etc.) make yield predictions for those nuclides unreliable by rather large factors. Secondly, observations of Be and B in metal-poor stars (Ryan et al. 1992, Duncan et al. 1992) showed that those elements behave as {\it primaries}, i.e. the Be/Fe and B/Fe ratios remain constant with metallicity. A {\it secondary } behaviour was {\it a priori} expected (since the yields of those nuclides, being proportional to spallated CNO, should increase with time/metallicity), and this expectation is not modified even by invoking extreme features for the GCR propagation in the early Galaxy (Prantzos et al. 1993a). The B data may be fixed by its neutrino production in supernovae, but  for Be one has  to assume that the CNO content of cosmic rays does not increase with time/metallicity (Duncan et al. 1992, Prantzos et al. 1993b). That bold conjecture  is, in fact, the only possibility, as shown by Ramaty et al. (1997) on the grounds of energetics: if the CNO content of GCR were lower in the past, much more energy in GCR would be required to compensate for that and to keep the Be/Fe ratio constant; but the required energy is much larger than available from supernovae explosions, which are the main energy source of GCR (Sec. 3.1 and Fig. 7a). The implications of that discovery for the (still debated) origin of GCR  are not clear yet (see Sec. 3.2 and text by Binns, this volume).

A new twist to the LiBeB saga came in the 2000s, with the discovery of $^6$Li in metal-poor halo stars (Asplund et al. 2006), after several unconvincing attempts in the 90ies. Be was expected to behave as secondary and found to behave as primary. $^6$Li was expected to behave as primary (since it is mostly produced by metallicity independent $\alpha + \alpha$ reactions in the early Galaxy, as argued by Steigman and Walker 1992) and found to display a "plateau", at a level $\sim$20 times lower than  the Spite plateau of Li. The $^6$Li plateau lies above the well-constrained contribution of standard GCR, calling for other explanations for its origin (Sec. 4.2 and Fig. 9).

In the following sections I discuss each one of the light nuclides (except for $^4$He), insisting on recent developments and current issues.\footnote{ For comprehensive reviews see: Reeves (1994), Prantzos et al. (1998) and Steigman (2006).}

\section{Deuterium and Helium-3}

The lives of D and $^3$He are intimately, but not totally, coupled: they are both produced in the Big Bang and D is rapidly turned into $^3$He inside stars. In the 80ies, a lot of effort was devoted to find how much of this $^3$He survived and was reejected in the ISM (e.g. Dearborn et al. 1986), in order to use the sum of D + $^3$He to constrain the baryonic density from SBBN (e.g. Yang et al. 1984, Walker et al. 1991). However, such attempts were futile, due to the (well known at the time) fact that stars can produce their own $^3$He (i.e. independently of any initial D), but also they can destroy D {\it and} $^3$He (i.e. without producing any $^3$He). In other terms, it is not {\it a priori} known whether the sum of D+$^3$He (used in such studies) has to stay constant, to decrease or to increase during galactic evolution. The evolution of the two nuclides should then be considered independently.

\subsection{Deuterium}

Modelling the Galactic chemical evolution (GCE) of deuterium is a most straightforward enterprise, since this fragile isotope is 100\% destroyed in stars of all masses
 and has no known source of substantial production other than BBN. If the boundary conditions of its evolution (namely the primordial abundance resulting from BBN and the present day one) were precisely known, the degree of astration, which depends on the adopted stellar initial mass function (IMF) and star formation rate, should be severely constrained.

\begin{figure*}
\centering
\includegraphics[clip=,angle=-90,width=0.95\textwidth]{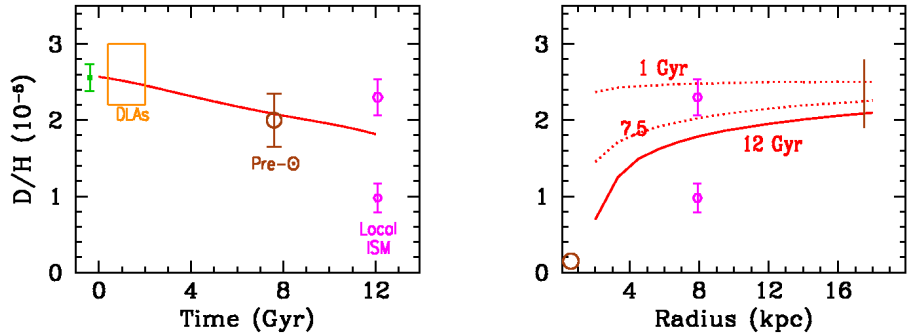}
\caption{{\it Left:} Evolution of deuterium in the solar neighborhood, as a function of time. The adopted   model satisfies all major local observational constraints, but is not unique (i.e. other satisfactory solutions may be found, where D is is slightly more destroyed, e.g. with a different IMF). Data are from Table 1. {\it Right:} Evolution of the deuterium abundance profile in the Milky Way disk; curves correspond to 1  Gyr,  7.5 Gyr (Sun's birth) and 12 Gyr (today), from top to bottom; the latter is to be compared to data for the present-day ISM. Data are from Table 1 for local values (at a radius of 8 kpc), from Rogers et al. (2005) at 16 kc and from Lubowich et al. (2000) in the inner Galaxy.} 
\label{Fig3}
\end{figure*}

The difficulty to determine the primordial D abundance in the 90ies pushed researchers to turn the problem upside down and try to determine that abundance through reasonable models of local GCE (assuming that the present day abundance is precisely known). Those efforts concluded that reasonable GCE models, reproducing the major observational constraints in the solar neihborhood, result only in moderate D depletion, by less than a factor of two (Prantzos 1996, Dearborn et al. 1996, Prantzos and Silk 1998, Tosi et al. 1998, Chiappini et al. 2002).

The primordial abundance of D is now well determined (Table 1), since observations of D in high redshift gas clouds agree with abundances derived from observations of the Cosmic Microwave Background combined to SBBN calculations; it points to a small D depletion up to solar system formation 4.5 Gyr ago (Fig. 3a). However, the present day abundance of D in the local ISM is now under debate. Indeed, UV measurements of the FUSE satellite  along various lines of sight suggest substantial differences (a factor of two to three) in D abundance between the Local Bubble and beyond it  (see Table 1 and Fig. 3a). Until the origin of that discrepancy is found (see Hebrard et al. 2005 and Linsky, this volume), the local GCE of D in the past few Gyr will remain poorly understood: naively, one may expect that a high value would imply strong late infall of primordial composition, while a low value would imply strong late astration (e.g. Geiss et al. 2002, Romano et al. 2006). In any case, corresponding models should also satisfy all other local observables, like the overall metallicity evolution and the G-dwarf metallicity distribution, which is not an easy task. It should be noted that the FUSE data may also be interpreted as suggesting an inhomogeneous composition for the local ISM, on scales of $\sim$500 pc  (for D, but also O and N, see e.g. Knauth et al. 2006), which does not seem to be the case for other elements (Cartledge et al. 2006).

\begin{figure}
\leavevmode
\begin{minipage}[b]{0.45\textwidth}
\includegraphics[clip=,width=2.45in]{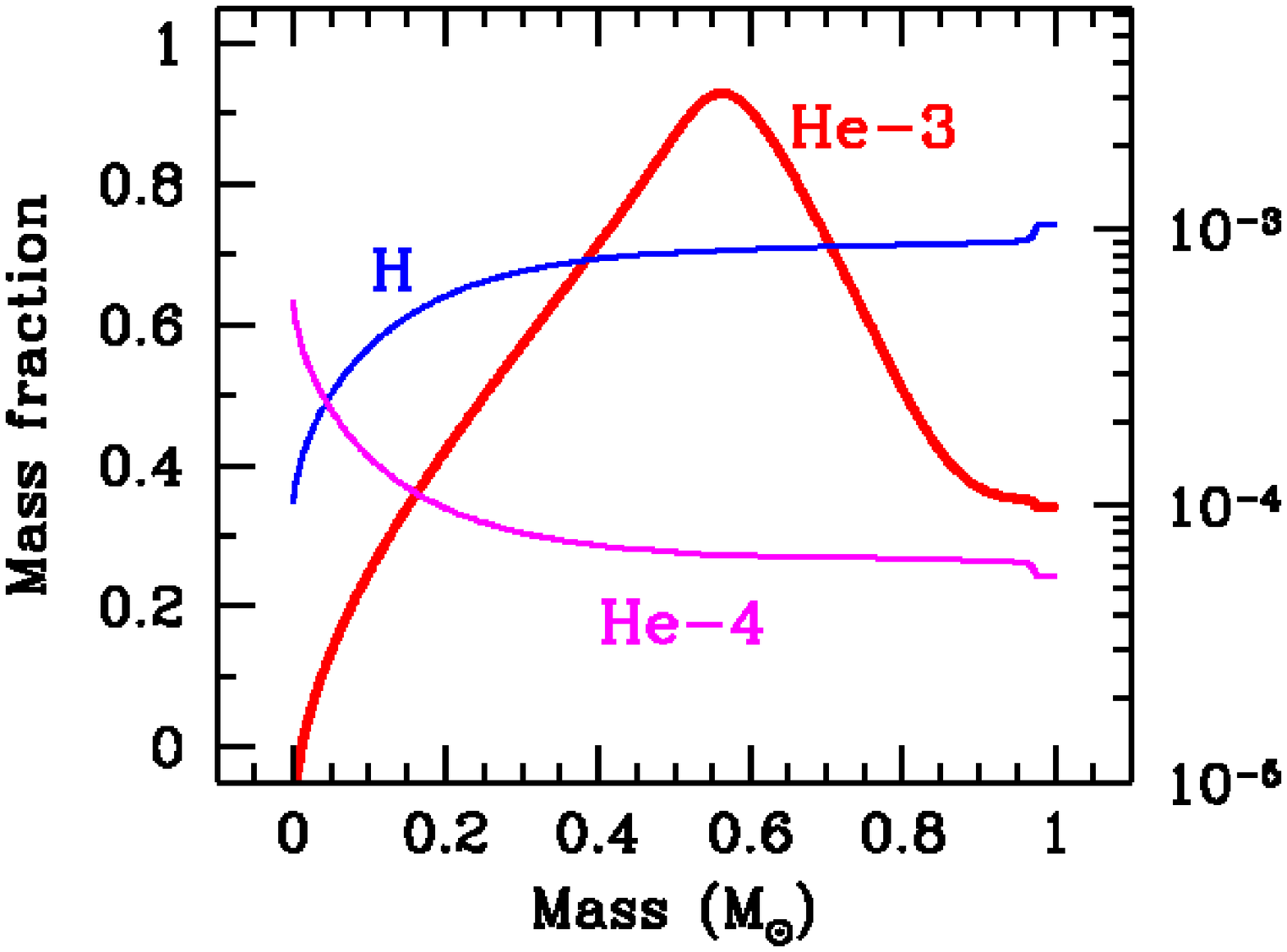}
\end{minipage}
\hfill
\begin{minipage}[b]{0.51\textwidth}
\includegraphics[width=0.98\textwidth]{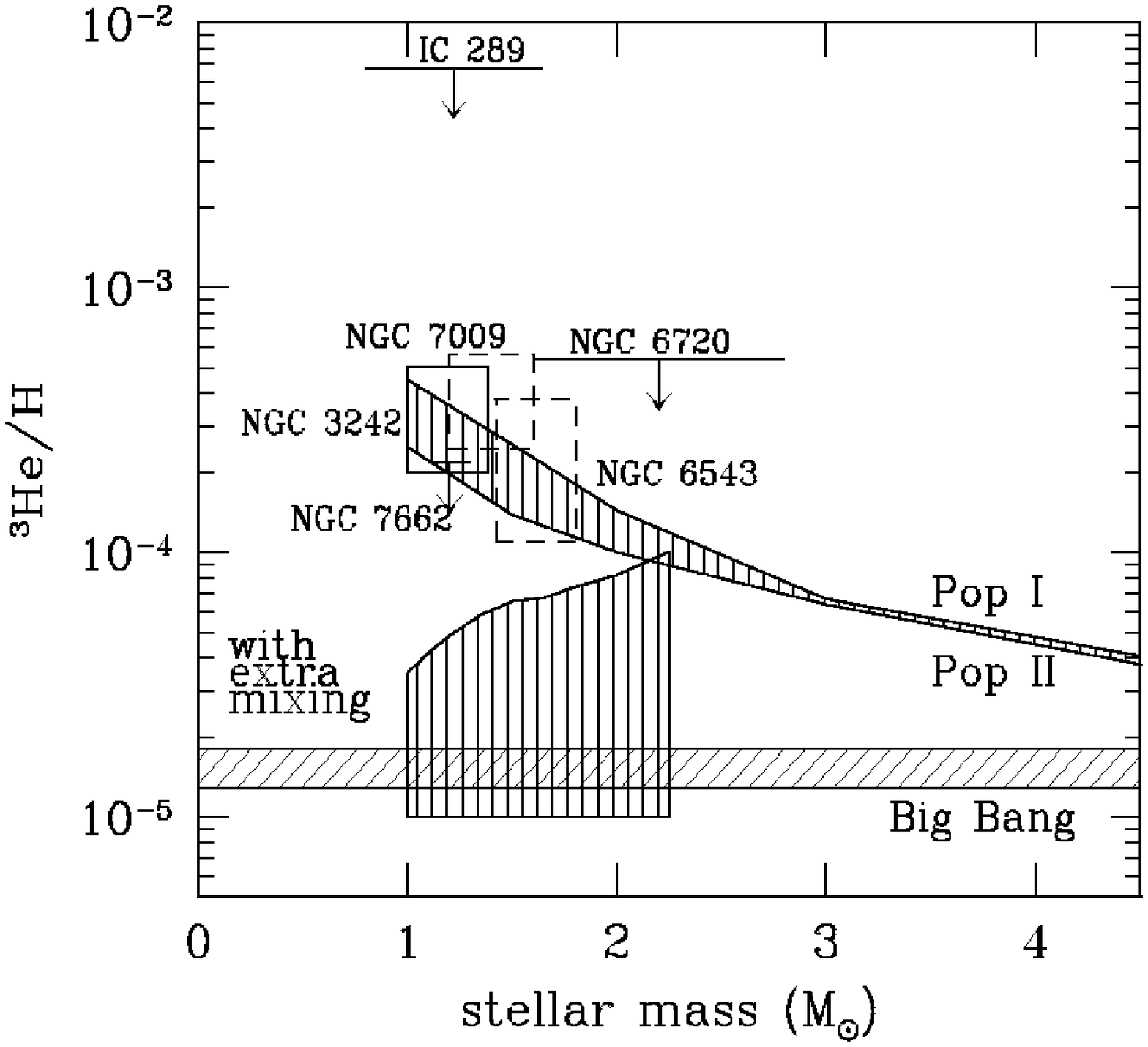}
\end{minipage}
\caption{{\it Left:} Abundance profiles of H, $^4$He (left axis) and $^3$He (right axis) in the present-day Sun, as a function of the mass coordinate; when low mass stars become red giants, the convective envelope reaches regions of enhanced $^3$He abundance and brings it to the surface. {\it Right:} Abundance of $^3$He in planetary nebulae (from Galli 2005). {\it Upper shaded aerea} indicates predictions of standard models, in agreement with observations ({\it within rectangles}); such high abundances lead to overproduction of $^3$He during galactic evolution (upper curves in Fig. 5). {\it Lower shaded aerea} indicates required level of production in order to avoid overproduction of $^3$He during galactic evolution (lower curves in Fig. 5); such a reduced yield may result from extra-mixing in red giants, also required on other observational grounds (Charbonnel 1995). It should affect 90-95\% of all stars below 2 \ms, while current observations of $^3$He in such stars would concern then the remaining 5-10\%. 
}
\label{Fig4}
\end{figure}

The evolution of D in the Galactic disk was considered originally with analytical models by Ostriker and Tinsley (1975), who found that D should be largely depleted in the inner disk.  Using numerical models (satisfying all the major observational constraints for the disk) Prantzos (1996) confirmed that finding (Fig 3b) and showed that the D/O profile of the disk offers a most sensitive test of its past history; unfortunately, such a profile has not been established in  realiable way yet.

\subsection{$^3$He}

Since the pioneering work of Iben (1967) stars are known to produce substantial amounts of $^3$He, through the action of p-p chains on the main sequence (see Fig. 4a). The net $^3$He yield varies steeply with mass (roughly $\propto M^{-2}$), since the p-p chains are less effective in more massive stars. In standard stellar models, 1-2 \ms \ stars are the most prolific producers.

Combining those yields with simple GCE models, Truran and Cameron (1971) and Rood et al. (1976) found that local abundances of $^3$He are  largely overproduced. Indeed, the current ISM abundance of $^3$He/H$\sim$1.-2 10$^{-5}$ is not very different from the pre-solar value (see Table 1 and Bania, this volume). In other terms, observations show that $^3$He abundance remained $\sim$constant through the ages, while standard stellar models combined to GCE models (e.g. Prantzos 1996, Dearborn et al. 1996, Galli et al. 1997, Romano et al. 2003) point to a large increase (Fig. 5, upper curves).

\begin{figure*}
\centering
\includegraphics[clip=,angle=-90,width=0.95\textwidth]{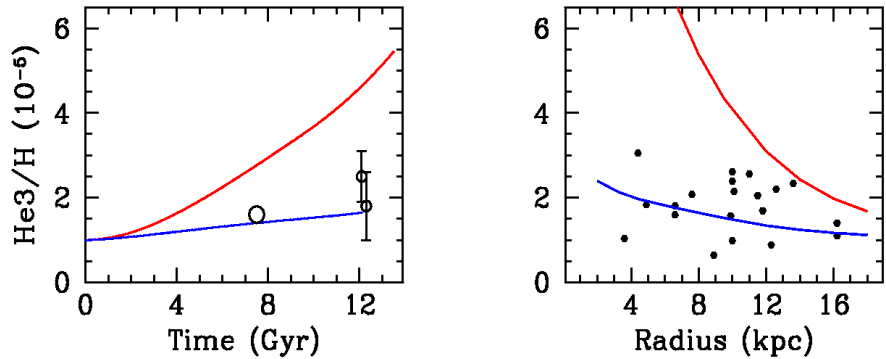}
\caption{Evolution of the abundance of $^3$He in the solar neighborhood
  as a function of time ({\it left}) and present day profile of $^3$He/H in 
  the Milky Way disk ({\it right}). In both cases, the {\it upper} curves are
  calculated with standard $^3$He yields from low mass  stars (and  clearly
  overproduce $^3$He) while the lower ones by assuming that 95\% of the $^3$He of low mass stars is destroyed by some non-standard mechanism; this latter assumption
  allows to satisfy observational constraints, but is not supported by the rare observations of $^3$He in planetary nebulae of presumably known mass (see Fig. 4b). In the {\it left panel} pre-solar $^3$He ({\it large circle}) and present day values in the local ISM ({\it small circles}) are from Table 1. In the {\it right panel}, ISM values are from Bania et al. (2002). 
} 
\label{jdrfig2}
\end{figure*}

A possible solution to the problem was suggested by Hogan (1995) and Charbonnel (1995). It postulates destruction of $^3$He in the red giant phase of Low mass stars through some "extra-mixing" mechanism, which brings $^3$He in H-burning zones. The "bonus" is a concommitant modification of the $^{12}$C/$^{13}$C isotopic ratio in red giants, in excellent agreement with observations (Charbonnel and do Nascimento 1998).

Thus, low and intermediate mass  stars should destroy in the red giant phase whatever $^3$He they produce on the main sequence. A possible drawback to the idea is that observations in (at least one) planetary nebulae of known  mass  are in  full agreement with standard model predictions, i.e. with no extra-mixing (see Fig. 4b and Galli 2005). GCE requires that in $>$90\% of the stars, $^3$He produced on the main sequence must be destroyed in the red giant phase, in order to avoid oveproduction (Fig. 5, lower curves). It may well be that current detections of $^3$He in planetary nebulae (see Bania, this volume) concern only the remaining $<$10\% of the stars, but it is still early to draw definitive conclusions.

\section{Beryllium and Boron}

The recent evolution of Be and B in the solar neighborhood  was considered already in Reeves and Meyer (1978). However, only in the 90ies spectroscopic observations of Be and B in metal poor halo stars of the MW became feasible, revealing an unexpected primary behaviour for those elements (Fig. 6a)\footnote{The case of boron is more complicated. As found in MAR, standard GCR (i.e standard equilibrium GCR spectra folded with well known spallation cross-sections of CNO nuclei) produce a $^{11}$B/$^{10}$B ratio of 2.4, instead of the pre-solar value of 4. Thus, 40\% of solar $^{11}$B should originate from another source, and this might well be $\nu$-nucleosynthesis in core collapse supernovae (Woosley et al. 1990, Olive et al. 1994); this produces primary B and could explain the observed linearity of B vs Fe (but not the one of Be).}.

\begin{figure*}
\centering
\includegraphics[clip=,angle=-90,width=0.95\textwidth]{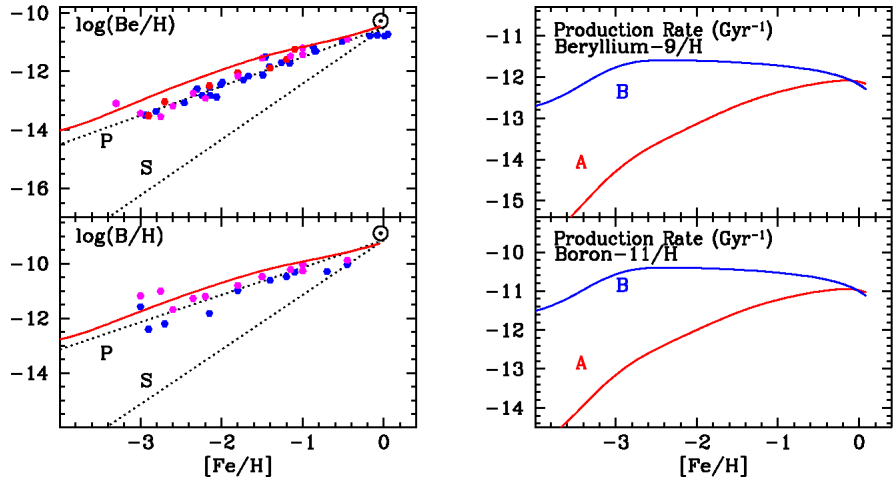}
\caption{ {\it Left:} Evolution of Be/H and B/H ({\it solid curves}) {\it assuming } that the GCR composition is independent of time (or ISM metallicity); Be and B are then produced as primaries, in agreement with observations. {\it Dotted curves} indicate primary (P) and secondary (S) behaviour with respect to Fe (while Be and B are produced from CNO, behaving not exactly as Fe). Note that $\sim$40\% of solar $^{11}$B has to be produced by a source other than standard GCR, like e.g. $\nu$-nucleosynthesis in supernovae, which is a primary process (this is not included in the figure). {\it Right:} Production rates of Be and B as a function of metallicity, for GCR components A (fast protons and alphas impinging on ISM CNO) and B (fast CNOs impinging on ISM H and He). Component A (producing secondary BeB) slightly dominates today, but component B ({\it assumed to be metallicity independent}) largely dominated during the halo phase, i.e. at [Fe/H]$<$-1, producing primary BeB. Standard GCR injection spectra, energetics and confinement are assumed throughout galactic evolution in this  figure.} 
\label{Fig6}
\end{figure*}

\subsection{Primary Be and B vs. energetics of GCR}

 As pointed out in MAR, there are two components in the production of BeB from GCR (Fig. 6b): component A  (fast protons and alphas impinging on ISM CNO) and component B (fast CNOs impinging on ISM H and He). Component A has, in principle, a production rate proportional to the steadily increasing CNO abundances of the ISM and produces secondary BeB. To boost the BeB production of component A at low metallicities, one may assume either that GCR (part of which is ``leaking'' out of the Galactic disk today) were much better confined in the early Galaxy (Prantzos et al. 1993a) or that their total energy content was much larger than today; in both cases, the number of induced reactions with ISM CNO (per unit H atom) is increased with respect to its current value. The former option was shown to be inefficient (although improving the situation, it cannot produce a linear BeB vs Fe relation, as found in Prantzos et al. 1993a) where the latter was shown to be unrealistic by Ramaty et al. (1997): already a large fraction (10-20\%) of the kinetic energy of SN goes into acceleration of GCR, and it is simply impossible to increase that fraction by a factor of, say, 100-1000 (as to compensate for the 100-1000 times lower CNO content of the ISM in the early Galaxy). This ``energetics barrier'' to the early production of Be is illustrated in Fig. 7a.

This leaves component B (presently sub-dominant) as the only option, at least for Be, which has no other known source. As suggested in Duncan et al. (1992) one has to {\it assume that the CNO content of GCR remained $\sim$constant since the earliest days of the Galaxy}. This bold assumption is tightly related to the, yet unsettled, question of the  GCR source composition (see Meyer et al. 1997 and references therein).

\begin{figure*}
\centering
\includegraphics[clip=,angle=-90,width=0.95\textwidth]{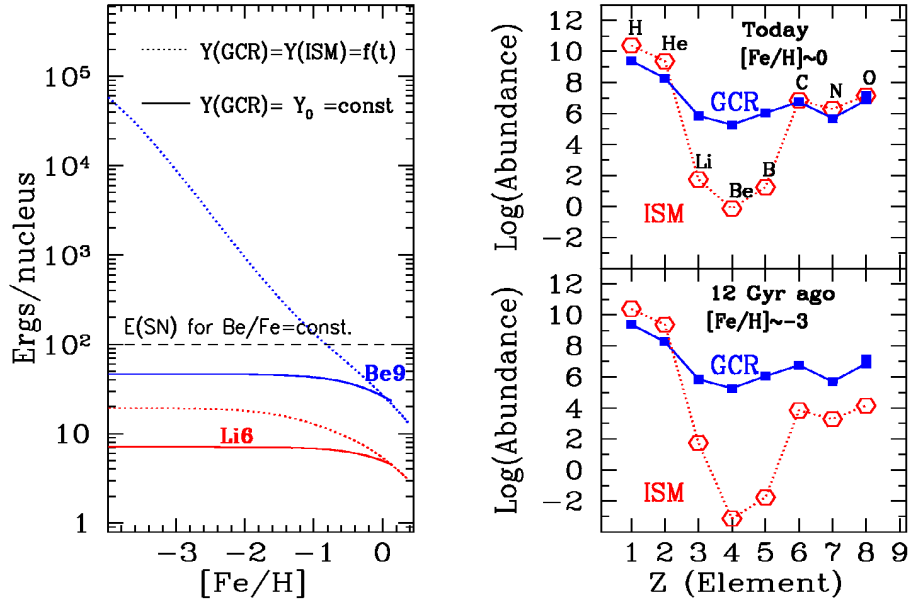}
\caption{{\it Left:}  Energy requirements (energy in accelerated particles per light nucleus produced) for the production of $^9$Be ({\it upper curves}) and $^6$Li ({\it lower curves}) through the interaction of GCR with the ISM, as a function of the ISM metallicity. $^6$Li is mostly produced by $\alpha +\alpha$ reactions and the energetics is quasi-independent of  metallicity.  This also happens in the case of $^9$Be, {\it if} it is assumed that the GCR composition is independent of time ({\it solid curves}); if the GCR composition scales with the one of the ISM ({\it dotted curves}), then extremely large energies are required for the production of $^9$Be below [Fe/H]$\sim$-1. Such energies are larger than available from a typical SN,  as indicated by the {\it horizontal dashed line}; it is assumed here that SN turn 20\% of their kinetic energy $E_{KIN}$=1.5 10$^{51}$ erg into GCR,  produce 0.1 M$_{\odot}$ of Fe and  induce the production of  10$^{-7}$ M$_{\odot}$ of Be, as to keep the Be/Fe ratio always constant (in agreement with observations). A time (or metallicity) dependent GCR composition should then  be excluded. {\it Right:} Composition of GCR ({\it solid curves, filled symbols}) and ISM ({\it dotted curves, open symbols}), today ({\it upper panel}, observed) and in the early Galaxy ({\it lower panel}, inferred); GCR {\it had to be much more metallic than the ambient ISM} in those early days, to account for the observed primary behaviour of Be.
} 
\label{Fig7}
\end{figure*}

\subsection{On the GCR source composition}

A $\sim$constant abundance of C and O in GCR 
can naturally be understood if SN
accelerate their own ejecta. However, the absence of unstable $^{59}$Ni 
(decaying through e$^-$-capture
within 10$^5$ yr) from observed GCR suggests that acceleration occurs 
$>$10$^5$ yr after the explosion (Wiedenbeck et al. 1999), 
when SN ejecta are presumably 
diluted in the ISM.  Obviously then, SN do not accelerate their own ejecta. 
However,  they can certainly accelerate the ejecta of their neigbours. Higdon
et al. (1998) suggested  that this happens in {\it superbubbles} 
(SB), enriched by the ejecta of many SN as to have a large 
and $\sim$constant metallicity. Since then, this became {\it by default}, 
the  "standard" scenario for the production of primary Be and B by GCR,
invoked in almost every  work on that topic.

However, the SB scenario suffers from several problems.
First,  core collapse SN are observationally associated
to HII regions (van Dyk et al. 1996) and it is well known that the
metallicity of HII regions reflects the one of the {\it ambient ISM} (i.e.
it can be very low, as in IZw18) rather than the one of SN.
Secondly, the scenario requires that SB in the early Galaxy retain most of their metals,
to the point of being much more metallic than the ambient ISM (say, by  factor of $\sim$1000, see 
Fig. 7b). But this goes against current wisdom: indeed, observations and hydrodynamical 
simulations suggest that small galactic units (such as those that merged to form the Galactic halo in the framework
of the hierarchical merging scenario) are metal poor, possibly because their weak gravity cannot retain the hot, metal-rich
ejecta of supernovae. It is hard to understand then why SB in the galactic sub-units forming the early Galaxy  would be so much more metal-rich than their environment (and, on top of that, with {\it always the same} quasi-solar metallicity).

Finally, Higdon et al. (1998) evaluated the time interval $\Delta t$ between
SN explosions in a SB to a 
comfortable $\Delta t \sim$3 10$^5$ yr, leaving enough time to $^{59}$Ni 
to decay before the next SN explosion and subsequent acceleration. 
However,  SB are constantly powered not only by SN
but also by the strong  {\it winds of
massive stars} (with integrated energy and acceleration
efficiency similar to the SN one, e.g. Parizot et al. 2004), 
which should  continuously
accelerate $^{59}$Ni, as soon as it is ejected from SN explosions.$^{59}$Ni should then be observed in GCR, which is not the case (Prantzos 2005). 
Thus, SB suffer exactly from the same problem that plagued SN as accelerators
of their own, metal rich, ejecta. Note that a loophole in the latter 
argument is suggested by Binns (this volume): only the
most massive stars ($>$30 \ms) display strong winds; such stars live for less than 6 Myr, i.e. less than the first 1/4 of the SB lifetime,
implying that the largest fraction of GCR ({\it if} accelerated in SB) should be free of $^{59}$Ni, in agreement with observations. It remains to be seen whether the
argument holds quantitatively, but even in that case the first two objections against the SB idea still hold.

The problem of the source and acceleration site of GCR, so crucial for the
observed linearity of Be and B vs Fe (but also for our understanding of GCR in general)  has not found a satisfactory explanation yet (at least to the opinion of the author of this paper).

\section{The Li isotopes}

The isotope $^7$Li holds a unique position among the $\sim$315 naturally occuring nuclides, since it is produced
by more than two nucleosynthesis sites: the hot early Universe, galactic cosmic rays (not only by $p+CNO$ but also by $\alpha+\alpha$ reactions), AGB stars, novae, and $\nu$-nucleosynthesis in core collapse SN. The contribution of the first two processes is relatively well known, while the remaining ones are hard to quantify at present (see e.g. Romano et al. 2003, Travaglio et al. 2001, for such attempts).

\subsection{From primordial $^7$Li to the "`Spite plateau"'}

\begin{figure}
\leavevmode
\begin{minipage}[b]{0.48\textwidth}
\includegraphics[clip=,angle=-90,width=0.98\textwidth]{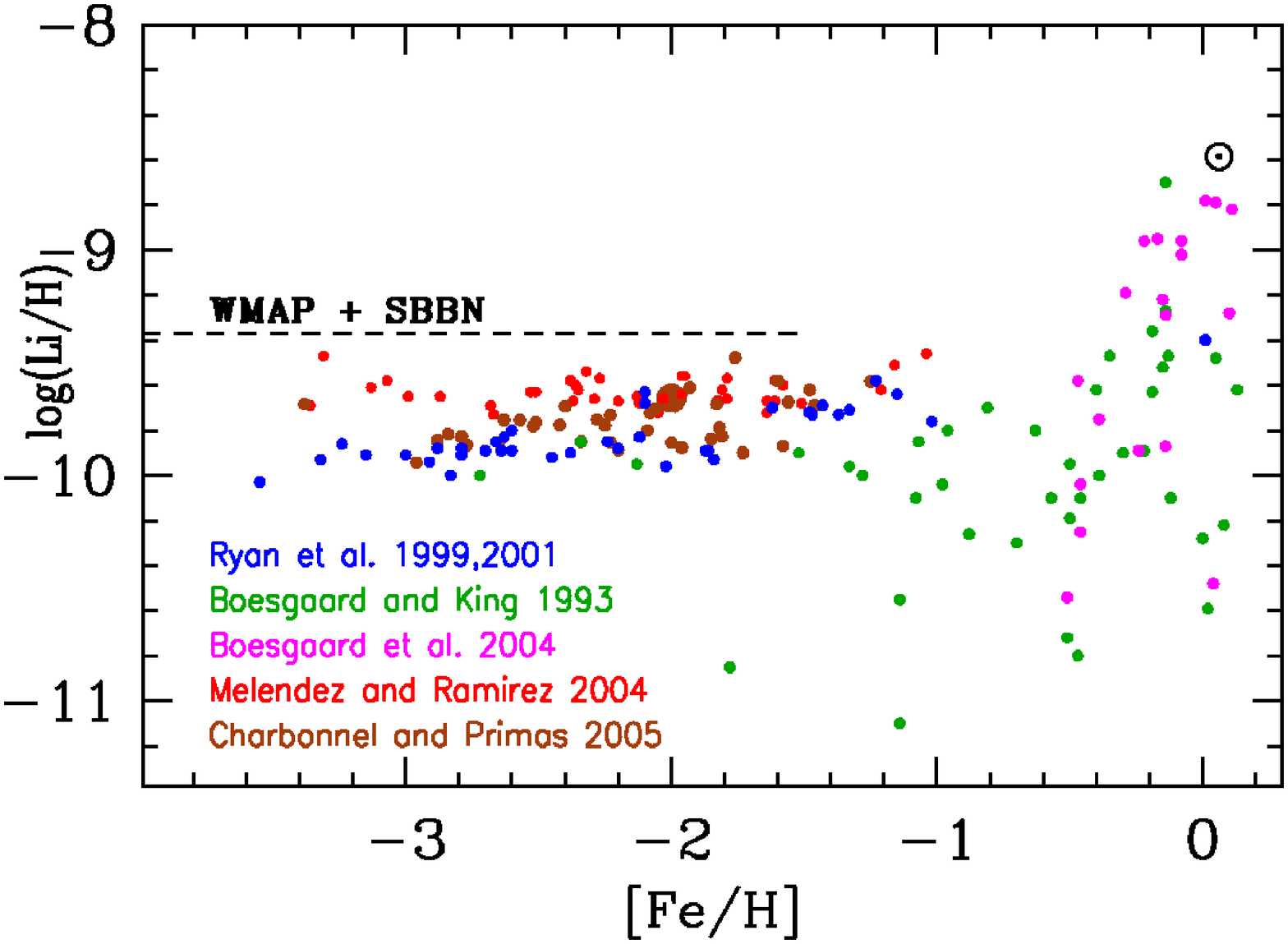}
\end{minipage}
\hfill
\begin{minipage}[b]{0.48\textwidth}
\includegraphics[clip, angle=-90,width=0.98\textwidth]{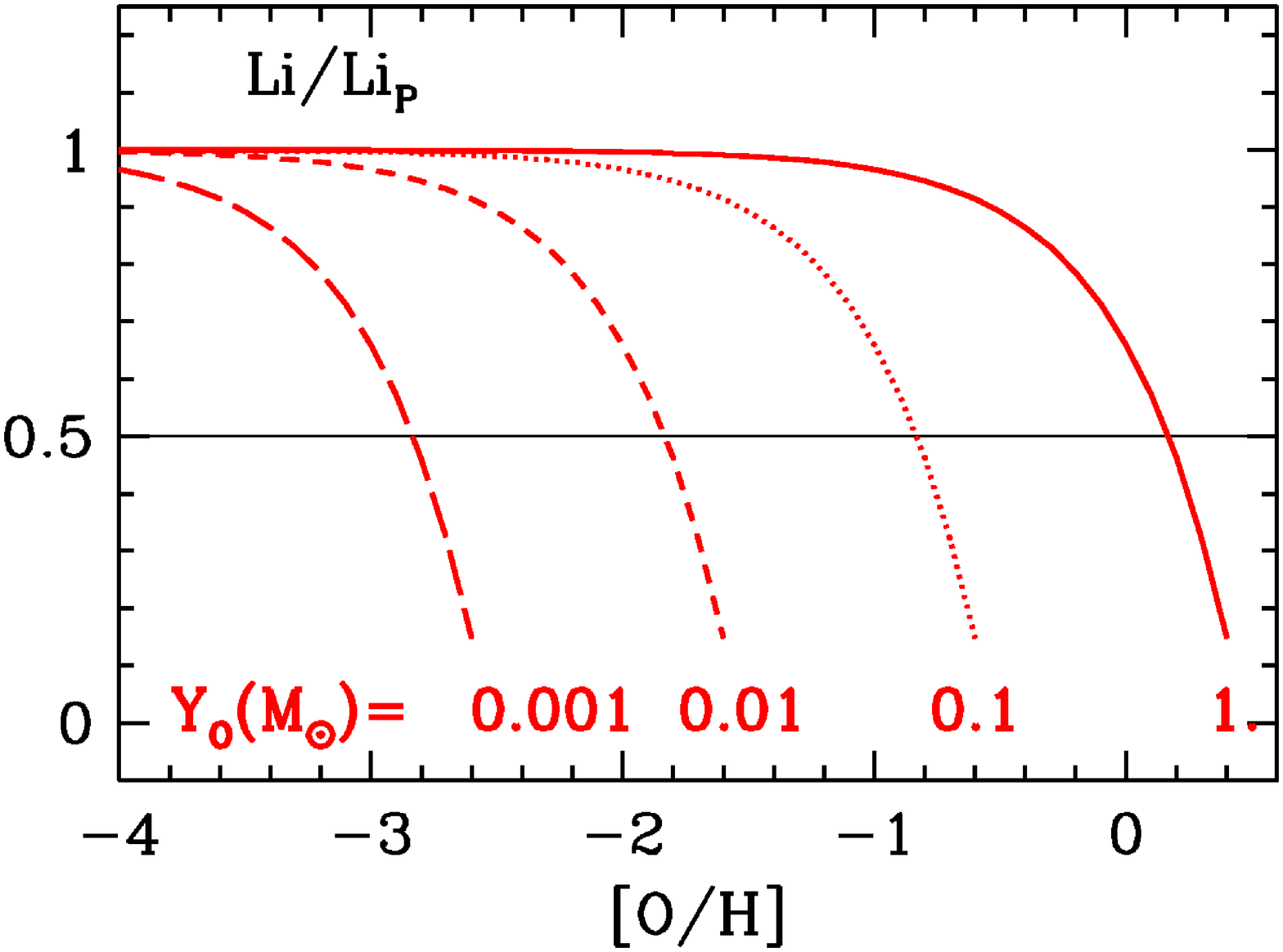}
\end{minipage}
\caption{ {\it Left:} Observations of Li in halo and disk stars of the Milky Way. The primordial Li value, obtained from the baryonic density of WMAP and calculations of SBBN, is indicated by a {\it horizontal dashed line}. The observed Li plateau at low metallicities depends sensitively on assumed stellar temperature, and differs from the WMAP value by factors 2-3. {\it Right:} Illustration of the problem encountered by the idea that the discrepancy is due to astration of Li by a Pop. III composed of exclusively massive stars, in the 10-40 \ms \ range (Piau et al. 2006): such stars necessarily eject metals, either through their winds (e.g. Nitrogen, in case of rotating stars) or through the final supernova explosion (e.g. Oxygen). Mixing the ejecta with various proportions of primordial material would result in Li depletion (by a factor of $\sim$2 in case of a 50-50 mixture), but the resulting metallicity would be much higher than the one of halo  stars, unless abnormally low oxygen yields were assumed (i.e. the curve in Fig. 8b parametrized with a yield $Y_O$=0.001 \ms, see text). 
}
\label{Fig4}
\end{figure}

The Li abundance of the "Spite plateau" (Li/H$\sim$ 1.-2 10$^{-10}\sim$const. for halo stars, down to the lowest metallicities) is a factor of 2-3 lower than the WMAP+SBBN value (Table 1 and Fig. 8a). Barring systematic errors (see Gratton, this volume), the conclusion is that primordial Li has been depleted, either (a) {\it before} getting into the stars it is observed today, or (b) {\it during} the lifetime of those stars. Two "depletion agents" have been proposed in the former case: decaying supersymmetric particles (Jedamzik 2004) and astration in a first generation of exclusively massive (mass range $m_*$=10-40 \ms) Pop. III stars (Piau et al. 2006). The latter idea,
however, suffers from a serious flaw, since in that case the metallicity of the ISM (out of which  the next stellar generation would form with depleted Li) would rise to levels much higher than those observed in EMP stars. This can be seen as follows (Prantzos 2006b):

Assuming that the current halo stellar mass ($M_H$=2 10$^9$ \ms) was initially in the form of gas, a fraction $f$ of which was astrated through massive stars,  the resulting Li mass fraction is $X_{Li}$=$(1-f) X_{Li,P}$ where $X_{Li,P}$ is the primordial Li abundance (assuming a return fraction $R\sim$1 for the astrating and metal producing stars). Similarly, the resulting oxygen abundance would be $X_O$=$m_O$/$M_H$, where the mass of oxygen $m_O=N_{SN} \ Y_O$ is produced by a number of supernovae $N_{SN}$= $f M_H/ m_*$, each one with a typical oxygen yield of $Y_O$ (in \ms, to be discussed below). Then: 
$${{X_{Li}}\over{X_{Li,P}}} \ = \ 1 - 0.28 {{(X_O/0.007) (m_*/40 M_{\odot})}\over{Y_O}} $$
That relation appears in Fig. 8 (right panel) as a function of log($X_O/X_{O,\odot}$), with adopted solar abundance $X_{O,\odot}$=0.007. The four  curves correspond to different assumptions about the typical oxygen yield of a massive star of Z=0, ranging from 0.001 to 1 \ms; only the first of those yields leads to large Li astration at low metallicities, but (as discussed in Prantzos 2006b), stellar models produce  generically more than 1 \ms \ of oxygen per massive star.
Another way to eject astrated material by Z$\sim$0 massive stars is through stellar winds, which require rapidly rotating stars (radiative pressure being inneficient at low  metallicities);  but rotating massive stars produce large amounts of nitrogen (which may in fact help explaining the observed primary-like N in EMP stars, e.g. Meynet et al. 2006), thus the problem of metal overproduction is not avoided in that case either.
Astration in massive Pop. III stars cannot solve the Li discrepancy between the Spite plateau and WMAP+SBBN \footnote{At least, not the 10-40 \ms stars suggested in Piau et al. 2006 (provided that current nucleosynthesis models for such stars are correct); 100 \ms \ stars collapsing to black holes would be better candidates (provided they eject a substantial fraction of their astrated envelope only, but not of their metal-rich core).}: even a small Li depletion should be accompanied by excessive  metal enhancement. 

Several mechanisms were proposed over the years to account for case (b) above, i.e. depletion {\it during} the stellar evolution within the stellar envelope: rotational mixing, gravity waves, microscopic diffusion etc. (e.g. Charbonnel and Primas and references therein). The main difficulty is to obtain a uniform Li depletion of $\sim$0.3 dex over the whole metallicity range of the plateau, with negligible dispersion. Richard et al. (2005) proposed a model with a few ingredients (microscopic diffusion coupled to levitation due to radiation pressure, and moderated by turbulent diffusion at the base of the convective envelope) which reproduces satisfactorily that feature. Such models are supported by recent
spectroscopic observations of stars in the metal-poor globular cluster NGC6397,  revealing trends of atmospheric abundance with evolutionary stage for various elements (Korn et al. 2006).

\begin{figure*}
\centering
\includegraphics[clip=,angle=0,width=0.96\textwidth]{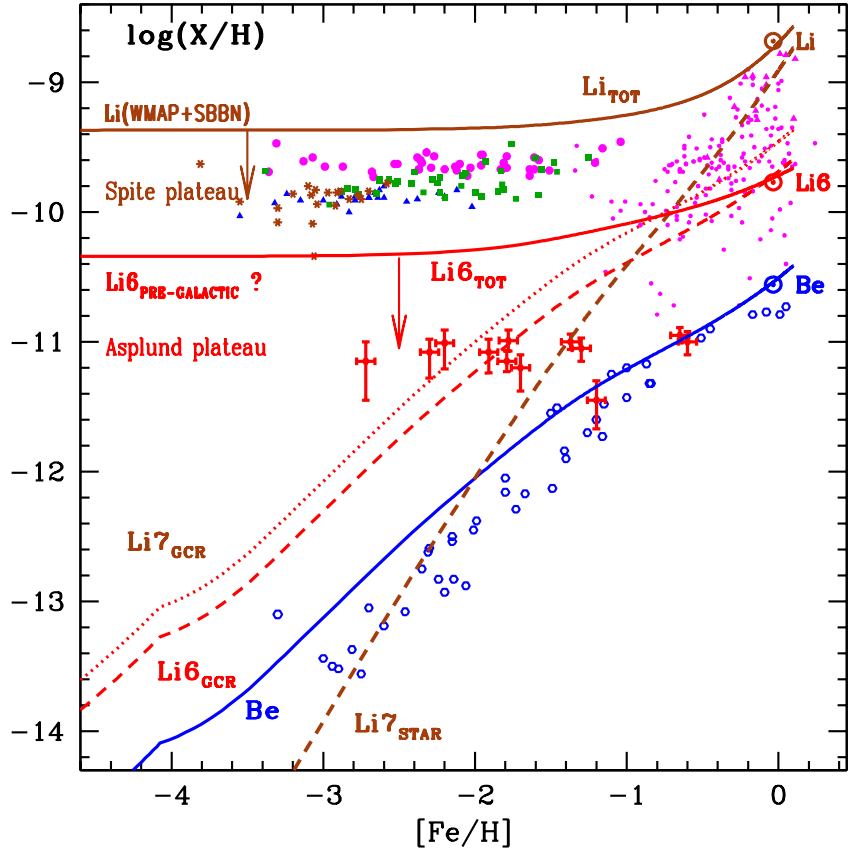}
\caption{Evolution of $^7$Li, $^6$Li and Be: observations vs. (a possible) theoretical picture. Observations are for total Li ({\it upper set of filled symbols}, the  "Spite plateau"), $^6$Li ({\it points with error bars} in the middle of the diagram, the "Asplund plateau" - to be confirmed) and Be ({\it lower set of open symbols}). The Spite plateau lies below the primordial value inferred from WMAP ({\it early part of upper solid curve}), most probably due to depletion of Li inside the stellar envelopes by $\sim$0.3 dex ({\it vertical arrow}). The more fragile $^6$Li should then suffer from even larger depletion ({\it vertical arrow}), its true value lying perhaps $\sim$0.7 dex above the Asplund plateau ({\it solid curve labeled} Li6$_{\rm TOT}$). The contribution of GCR to $^6$Li ({\it dashed curve labeled } Li6$_{\rm GCR}$) is well constrained once the evolution of Be ({\it solid curve labeled} Be) is reproduced; it can explain pre-solar $^6$Li (as already found in MAR), but it is clearly insufficient to explain the ``Asplund plateau'' and even less so the "undepleted" $^6$Li plateau. The GCR component of $^7$Li ({\it dotted curve labeled } Li7$_{\rm GCR}$) is also well defined and contributes marginally to total Li. Finally, the stellar (AGB or nova) $^7$Li component ({\it dotted curve labeled} Li7$_{\rm STAR}$) is required to explain $\sim$60\% of pre-solar Li; note that only the late part of that component (at [Fe/H]$>$--0.7) is constrained, by the upper envelope of the Li data (Note: 
Li$_{\rm TOT}$=Li7$_{\rm TOT}$+Li6$_{\rm TOT}$ and Li7$_{\rm TOT}$=Li7$_{\rm BBN}$+Li7$_{\rm GCR}$+Li7$_{\rm STAR}$.)} 
\label{Fig10}
\end{figure*}

\subsection {Early $^6$Li: primordial, pre-galactic or simply stellar ?}

The surprising detection of a $^6$Li "plateau" in metal-poor stars of the Galaxy (Asplund et al. 2006) challenges our uderstanding of the origin of light nuclides. The value of the "Asplund plateau",
$^6$Li/H$\sim$10$^{-11}$, is $\sim$10 times higher than the one contributed from standard GCR (accounting for the observed evolution of Be) at [Fe/H]=--3 (see Fig. 9), and $\sim$1000 times larger than resulting from SBBN in the early Universe. The discrepancy is even more serious if Li isotopes are depleted in stellar atmospheres (see Sec. 4.1), since $^6$Li is more fragile than $^7$Li; its true abundance could be then (considerably) higher than  10$^{-11}$.

Several ideas have been put forward to explain the production of a high early abundance of $^6$Li:

- (a) primordial, in non-standard Big Bang nucleosynthesis involving the decay/annihilation of massive particle (Jedamzik 2004). The bonus of that idea is that such particles could also destroy part of the primordial $^7$Li (releasing the tension between WMAP+SBBN and observations) but this is less appealing now, since observations favour stellar depletion (Sec. 4.1).

- (b) pre-galactic, by fusion of $\alpha$-particles; these could be  accelerated by the energy released i) during structure formation (Suzuki and Inoue 2002) or ii) from accretion onto super-massive black holes or (iii) from an early generation of Pop. III masive stars (Reeves 2005 and this volume). A critique of those ideas, based on a careful evaluation of the energetics of $^6$Li production from energetic particles (see Fig. 7a) is made in Prantzos (2006a); the first two appear much less promising than the last one.

- (c) stellar, by {\it in situ} reactions of energetic particles (mostly $^3$He+$^4$He and assuming an enhanced $^3$He abundance)  in the atmosphere of the stars during the 10 Gyr of their evolution (Tatischeff and Thibaud 2006). The stars are required to be very active in accelerating particles ($\sim$10$^6$ times the activity of the present-day Sun in their early main sequence), that activity being atributed to their rapid rotation. The values of  the ``Asplund plateau'' can then be reproduced in some extreme cases.

 The early $^6$Li plateau is the latest (but probably not the last) twist in the saga of the light nuclides. More data and a thorough understanding of the stellar properties are required before concluding whether the answer to the puzzle lies among (a), (b) or (c) above, or it is something completely different.

\section{Summary}

The {\it x-process}  turned out to be the most complex of all the nucleosynthetic processes envisioned in
B$^2$FH. Despite 50 years of progress in theory and observation, it is still unknown where most of $^3$He and $^7$Li and a large fraction of $^{11}$B come from. The origin of early $^6$Li remains equally mysterious, while the degree of astration of D in the solar neighborhood is poorly known. This is certainly good news: we shall have exciting things to discuss for Johannes' 90ieth anniversary! 

\bigskip
\noindent
{\bf Acknowledgements } I am grateful to the organizers for their kind invitation and for giving me the opportunity to participate in such an interesting meeting, celebrating the contributions of Johannes Geiss to our understanding of the origin of the light nuclides. {\it Bon anniversaire} Johannes !


\end{article}

\end{document}